\documentclass[twocolumn,showpacs,preprintnumbers]{revtex4}

\input{epsf}

\def\APP{{\it Acta Phys. Pol.} }

\def\IJMP{{\it Int. J. Mod. Phys.} }

\def\JP{{\it J. Phys.} }

\def\NP{{\it Nucl. Phys.} }
\def\PL{{\it Phys. Lett.} }
\def\PR{{\it Phys. Rev.} }
\def\PRL{{\it Phys. Rev. Lett.} }
\def\PRTS{{\it Physics Reports} }

\def\RPP{{\it Rep. Prog. Phys.} }

\def\SP{{\it Soviet. Phys. JETP} }

\def\ZP{{\it Z. Phys.} }

\def\al{\alpha}
\def\be{\beta}
\def\ga{\gamma}
\def\de{\delta}
\def\ep{\epsilon}

\def\ka{\kappa}
\def\la{\lambda}

\def\ta{\tau}

\def\ph{\phi}

\def\om{\omega}

\def\De{\Delta}

\def\La{\Lambda}

\def\Ph{\Phi}

 \def\frac#1#2{{\textstyle{{#1}\over
{#2}}}} 
\def\lsim{\mathrel{\rlap{\lower4pt\hbox{\hskip1pt$\sim$}}
    \raise1pt\hbox{$<$}}} \def\gsim{\mathrel{\rlap{\lower4pt\hbox{\hskip1pt$\sim$}}
    \raise1pt\hbox{$>$}}}
\def\sqr#1#2{{\vcenter{\vbox{\hrule height.#2pt
         \hbox{\vrule width.#2pt height#1pt \kern#1pt
         \vrule width.#2pt}
         \hrule height.#2pt}}}} 

\def\beq{\begin{equation}}
\def\eeq{\end{equation}}
\def\beqa{\begin{eqnarray}}
\def\eeqa{\end{eqnarray}}

\begin{document}

\title{Topological Defect Densities in Type-I Superconducting Phase Transitions}

\vskip 0.2cm

\author{J. P\'aramos$^1$, O. Bertolami$^1$, T.A. Girard$^2$ and P. Valko$^3$}

\vskip 0.2cm

\affiliation{$^1$Instituto Superior T\'ecnico, Departamento de
F\'{\i}sica, \\ Av. Rovisco Pais 1, 1049-001 Lisboa, Portugal}

\vskip 0.2cm

\affiliation{$^2$Centro de F\'{\i}sica Nuclear, Universidade de
Lisboa, \\ Av. Prof. Gama Pinto 2, 1649-003, Lisboa, Portugal}

\vskip 0.2cm

\affiliation{$^3$Department of Physics, Slovak Technical University,
\\ Ilkovicova 3, 812-19 Bratislava, Slovakia}

\vskip 0.2cm

\affiliation{E-mail addresses: x\_jorge@netcabo.pt;
orfeu@cosmos.ist.utl.pt; criodets@cii.fc.ul.pt;
valko@elf.stuba.sk}

\vskip 0.5cm

\date{\today}

\begin{abstract}
We examine the consequences of a cubic term addition to the
mean-field potential of Ginzburg-Landau theory to describe first
order superconductive phase transitions. Constraints on its existence are obtained
from experiment, which are used to assess its impact on
topological defect creation. We find no fundamental changes in
either the Kibble-Zurek or Hindmarsh-Rajantie predictions.

 \vskip 0.5cm

\end{abstract}

\pacs{74.20.De, 74.55.+h, 11.27.+d  \hspace{2cm}Preprint DF/IST-1.2002}

\maketitle

\section{Introduction}

It is generally believed that the Universe, evolving from the
initial Big Bang, underwent a series of symmetry-breaking phase
transitions \cite{Kibble, Linde} accompanied by the creation of
topological defects, frustrations of the unbroken phase within the
broken one, induced by continuity of the order parameter values.
These defects appear as magnetic monopoles, cosmic strings, domain
walls and textures.

Direct experimental tests of these ideas are unfeasible, but
transitions described by similar equations occur in experimentally
accessible condensed matter systems, and a new trend has unfolded
which compares these two systems. This ``cosmology in the
laboratory'' relies on the fact that the dynamics of phase
transitions lie in universality classes and the cosmological ones
are hence analogous to those of condensed matter. For instance,
vortices created in the superfluid phase transitions of $^4He$ and
$^3He$ have been studied experimentally (see e.g. Ref.
\cite{Bunkov, Vollhardt} for extensive discussions) following an
earlier suggestion in which common features with cosmic strings
have been noted \cite{Zurek}. Similarities between cosmological
phase transitions and the isotropic-nematic phase transition in
liquid crystals were studied in Refs. \cite{Chuang,Bowick}.
Analogy with the thermodynamics and transitions in polymer chains
was drawn in Ref. \cite{Bento}.

The case of superconductors is of particular interest as the
associated phase transition involves a local gauge
symmetry-breaking process.

In superconductors, cosmic strings manifest themselves as flux
tubes or vortices. Experiments aimed to observe defect densities
in high-$T_c$ materials \cite{Carmi} have lead to contradictory
results with respect to the density predicted by the Kibble-Zurek
(K-Z) mechanism \cite{Kibble}. This prediction is however accurate
only for global gauge symmetry breaking, a situation where the
geodesic rule for phase angle summation is valid. A local gauge
treatment by Hindmarsh-Rajantie (H-R) \cite{Hindmarsh-Rajantie}
identifies a new mechanism for defect generation, which leads to a
prediction well below the first Carmi-Polturak experiment
sensitivity, although the prediction for the second one is in
reasonable agreement with observation.

The above experiments were both conducted in type-II materials,
which exhibit a second order phase transition. The question
naturally arises as to the extent of changes in the defect density
predictions for type-I superconductors. This is the motivation for
our work.

Distinction between type-I and type-II superconductors is
traditionally made through the Ginzburg-Landau (G-L) parameter $\ka
= \la /\xi$, the ratio between the magnetic field penetration
length, $\la$, and the order parameter (scalar field) coherence
length, $\xi$. These characteristic length scales are obtained, in
the presence of a gauge field $\vec{A}$, from the free energy
density

\beq 
F(\Ph) = {1 \over 2m_e} \left|i \hbar \vec{\nabla} \Ph - {e
\over c} \vec{A} \Ph \right|^2 + V(\Ph) + {1 \over2} \vec{\mu}
\cdot (\vec{\nabla} \times \vec{A})~~, 
\label{free} 
\eeq

\noindent where $\vec{\mu}$ is the magnetic moment of the
specimen, $m_e$ is the electron mass and $\Ph$ is the order
parameter. The G-L potential is usually written as \cite{Zurek}

\beq 
V(\Ph) = \al \Ph^2 + {\be \over 2} \Ph^4~~, 
\label{pot} 
\eeq

\noindent where $\al$ is assumed to depend linearly on the
temperature, $\al=\al'(t-1)$, $t \equiv T/T_c$, $\al'$ and $\be$
are constants, and $T_c$ is the critical temperature. Thus one
obtains

\beq
\la = \sqrt{{ m_e c^2 \over 4 \pi e^2}{\be \over |\al|}}~~,
\label{lambda}
\eeq

\noindent and

\beq 
\xi = {\hbar \over \sqrt{2m_e |\al|}}~~. 
\label{xi} 
\eeq

\noindent

At $T=0$ the coherence length is given by $\xi_0 = \hbar /
\sqrt{2m_e \al'}$, with $ \ka \sim \sqrt{\beta}$. For $\ka > 1 /
\sqrt{2} $, the transition is second order, and $\xi_0$ is
typically less than $\sim 0.04 ~ \mu m $; for
 $\ka < 1 / \sqrt{2} $, the transition is first order, with $\xi_0$ typically
 greater than $\sim 0.08 ~ \mu m$.
In general, both are second order for $H = 0$. First order
transitions arise from the external field term in Eq. (\ref{free})
in the event that a characteristic sample dimension is greater
than $\la$.

\begin{figure}
\epsfysize=4cm \epsffile{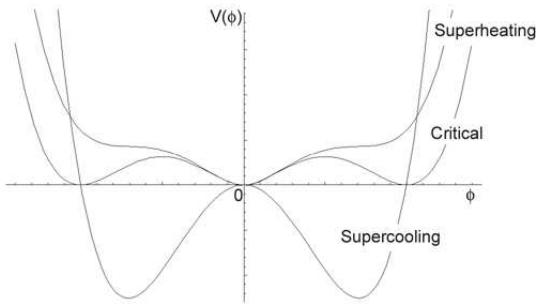} 
\caption{Characteristic
Potential curves.} 
\label{curves}
\end{figure}

In thermal field theory (TFT) a first order phase transition
arises from consideration of 1-loop radiative corrections to a
potential of the form of Eq. (\ref{pot}), which introduce a
barrier between minima through a cubic scalar field term, as

\beq 
V(\Ph) = \al \Ph^2 - \ga |\Ph|^3 + {\be\over2} \Ph^4~~,
\label{3pot} 
\eeq

\noindent where $\ga(T) = (\sqrt{2} / 4 \pi) e^3 T$
\cite{Hindmarsh}. As in the previous case, $\be$ is constant and
$\al=\al'(t-1)$ is linear with temperature.

A similar $ - \ga |\Ph|^3 $ term however, arises in considerations
of gauge field fluctuations in the normal-to-superconductor phase
transition \cite{Halperin, Hove}, with

\beq 
\ga = 8 \mu_0 {e \over \hbar c}\sqrt{\pi \mu_0} T_c ~~,
\label{fluct_gamma} 
\eeq

\noindent resulting in a first order phase transition for all
values of $\ka$. Crossovers between first and second order
transitions arise in considerations of thermal fluctuations
\cite{Hove}, as also in nonlocal BCS treatments \cite{Brandt}.

With this is mind, we adopt a potential of the form of Eq.
(\ref{3pot}) and explore constraints on $ \ga(T) $ from
experiment. The results are compared with TFT 1-loop radiative
corrections. Comparison with the results of Ref. \cite{Halperin},
which deals with temperatures close to $T_c$, is attained in the
limit $t \rightarrow 1 $.

These results are then used to analyze the impact of the cubic
term on the K-Z and the H-R predictions for type-I
superconductors, namely on the models themselves and on a possible
nucleation suppression due to the slowing down of the transition
induced by the potential barrier.

\section{Temperature sensitivity bound}

A first-order superconductive phase transition manifests
supercritical fields under variation of the temperature, as shown
in Figure \ref{curves}. The superheating curve is given by the
condition $ {dV \over d \Ph} = {d^2 V\over d \Ph^2} = 0 $, for a
certain value of $\Ph \neq 0$. This yelds $ \al = 9 \ga^2 / 16 \be
$.

In contrast, the supercooling curve is given by the condition $
{d^2 V(\Ph) \over d \Ph^2} = 0 $ for $\Ph=0$, corresponding to $
\al = 0 $. The unobservable critical curve is given by the
condition $ V(0) = V(\Ph_c) $ and ${dV(\Ph_c) \over d \Ph} = 0$,
where $\Ph_c$ is the non-vanishing  minimum of the potential. This
corresponds to $ \al = \ga^2 /2 \be $.

Introducing the linear dependencies $\al = \al' (t-1)$ and $
\ga(t) = \de ~ t$, we obtain for the superheating curve

\beq 
\al'(t-1) = {9 \over 16} {\de^2 \over \be}t^2~~,
\label{sh_cond} 
\eeq

\noindent and hence

\beq 
t_{sh} = {2 \over 1 + \sqrt{1 - {9 \over 4} {\de^2 \over \al'
\be}}} \sim 1 + {9 \over 16} { \de^2 \over \al' \be}~~.
\label{t_sh} 
\eeq

\noindent The superheating curve converges to a zero-field shift
in temperature from $T_c$ by $ (9 \de^2 / 16 \al' \be) T_c $. This
shift is undetected at the current experimental temperature
sensitivity of $ \De t_{exp} \sim 10^{-3}$ \cite{tempmeas}.
Therefore, a bound for the slope of $\ga$ is

\beq 
{9 \over 16} { \de^2 \over \al' \be} < \De t_{exp}~~.
\label{tbound} 
\eeq

The supercooling transition occurs at $ t = 1 $, the critical
temperature, since it is determined solely by $\al = 0 $ (this
neglects a small correction to $\al$, as given by Ref.
\cite{Halperin}).

\section{Superheating perturbation bound}

\begin{table}
\begin{ruledtabular}
\caption{Critical properties of  $Sn$ and $Al$} \label{table1}
\begin{tabular}{|c|c|c|c|c|}

    \hline

Material & $T_c ~(K)$ & $H_c(0)~(G) $ & $ \xi_0 ~(\mu m) $ & $ \la ~(nm) $ \\

\hline \hline

Sn & $ 3.7 $ & $ 309 $ & $ 0.23 $ & $ 34 $ \\

\hline

Al & $ 1.2 $ & $ 105 $ & $ 1.6 $ & $ 16 $ \\

\end{tabular}
\end{ruledtabular}
\end{table}

Measurements of the supercritical fields have commonly been
performed on microspheres of type-I materials as a means of
determining $\ka$. Table \ref{table1} indicates the critical
properties of two, Sn and Al. The existence of a cubic term should
also be manifest in the presence of  a magnetic field. To assess
its influence, we repeat Ref. \cite{Ginzburg} calculations,
including the cubic term in the potential. For a small
superconducting sphere of radius $a$, the magnetic moment is given
by \cite{Ginzburg}

\beq 
{\mu \over V} = -3 \left[1-{3\la \over a \Ph_0}coth{a \Ph_0
\over \la} + {3\la^2 \over a^2 \Ph_0^2}\right] {H\over8\pi}~~,
\label{mu} 
\eeq

\noindent where $\Ph_0 \equiv \Ph / \Ph_\infty$ and $\Ph_\infty^2
\equiv m_ e c^2 / 4\pi e^2 \la^2$.

After a somewhat lengthy computation (see Appendix), the reduced
superheating field, $h_{sh} \equiv H_{sh} / H_c$, is given by

\beq 
h_{sh} = \left(1 + {4 \over \sqrt[4]{15}} \ga_G \right)
h_{sh}^0~~, 
\label{h_sh_gamma}
\eeq

\noindent where $\ga_G \equiv 3 \ga / 2 \sqrt{|\al| \be}$ is a
dimensionless parameter, and $h_{sh}^0$ is the ``unperturbed"
($\ga=0$) superheating field.

Generally, such measurements have been obtained with colloids, and
the size distributions of the microspheres  \cite{data1, data2,
data4, data5, data6} provides a statistical error which renders a
direct fit of $\ga_G$ from $h_{sh}(t)$ data unfeasible. Although
the measurement reported in Ref. \cite{data3} was conducted using
single microspheres, non-local and impurity effects lead to a
large theoretical uncertainty in

\beq 
h_{sh}^0(t) = {1 \over \sqrt{\ka \sqrt{2}}} h_c^0(0) (1 -
t^2)~~, 
\label{parabola} 
\eeq

\noindent where $h_c^0(0)=H_c(0)/H_c(t)$, which is itself an
approximation valid only close to $T_c$ \cite{hcorr}.

Since the supercooling field implies the evaluation of a second
derivative at the origin, it can be easily seen that the presence
of a cubic term has no effect:

\beq 
{d^2\over d \Ph^2} (\ga_G \Ph^3) = 6 \ga_G \Ph \rightarrow
0~~. 
\label{noeffect} 
\eeq

The effect of $\ga$ (through $\ga_G$) on the superheating field
must be small, otherwise it would have been already detected;
therefore, we must have $ \ga_G \ll 1$, which is not valid for
relative temperatures in the range

\beq 
1 - {9 \over 4} {\de^2 \over \al' \be} < t < 1 ~~.
\label{interval} 
\eeq

\noindent For this interval to be vanishingly small,

\beq 
{9 \over 4} { \de^2 \over \al' \be} \ll 1 ~~. 
\label{hbound}
\eeq

\noindent This is a weaker bound than the one of Eq.
(\ref{tbound}).

For Al and Sn with maximum critical fields of order $10^2 ~G$, the
shift between $h_{sh}$ and $h_{sh}^0$ is less than $ 10^{-2} ~G $,
well below the sensitivity of measurements \cite{data1, data2,
data3, data4, data5, data6}.
For this reason, we simply drop the bound of Eq.
(\ref{hbound}) and consider only the one of Eq. (\ref{tbound}).
Conversely, a breakdown of the perturbation expansion of the
superheating reduced field would imply a superheating temperature
shifted to $ t_{sh} = 1.6$. Also notice that no ``spikes" should
be seen in the $ H - T $ superheating curve for values of $t$ in
the ``exclusion" interval as the field values are quite small
there.

\begin{table}
\begin{ruledtabular}
\caption{Derived quantities and bounds for $\de$} \label{table2}
\begin{tabular}{|c|c|c|}

Material & Sn & Al\\

\hline

$ \al' ~(J) $ & $ 1.15 \times 10^{-25} $ & $ 2.38 \times 10^{-27} $\\

\hline

$ \be ~(J.m^3) $ & $ 4.72 \times 10^{-54} $ & $ 2.16 \times 10^{-56} $\\

\hline \hline

$ \al' ~(eV^2) $ & $ 3.61 \times 10^{-1}  $ & $ 7.45 \times 10^{-3} $\\

\hline

$ \be $ & $ 9.45 \times 10^{-4} $ & $ 4.32 \times 10^{-6} $\\

\end{tabular}
\begin{tabular}{|c|c|c|c|c|}

bound & $ t_{sh} $ shift & $ \de (eV) $ & $ t_{sh} $ shift & $ \de (eV) $ \\

\hline \hline

$ h_{sh} $ & $ 0.25 $ & $ 1.23 \times 10^{-2} $ & $ 0.25  $ & $ 1.20 \times 10^{-4} $\\

\hline

$ \De T_{exp} $  & $ 10^{-3} $ & $ 7.78 \times 10^{-4} $ & $ 10^{-3}  $ & $ 7.57 \times 10^{-6} $\\

\hline \hline

Ref. \cite{Halperin} & $ 5.19 \times 10^{-9} $ & $ 1.77 \times 10^{-6} $ & $ 2.32 \times 10^{-6}  $ & $ 3.64 \times 10^{-7} $\\

\hline

TFT & $ 2.92 \times 10^{-9} $ & $ 1.33 \times 10^{-6} $ & $ 3.24 \times 10^{-6}  $ & $ 4.31 \times 10^{-7} $\\

\end{tabular}
\end{ruledtabular}
\end{table}

Table \ref{table2} provides a comparison of the bounds on $\de$
with the prediction of Ref. \cite{Halperin}. The analogy between
cosmology and condensed matter prompts for comparison with the TFT
cubic term also. To do this, we compute the associated slope of
$\ga(t)$ from $\ga(T) = (\sqrt{2} / 4 \pi) e^3 T $, obtaining $
\de = (\sqrt{2} / 4 \pi) e^3 T_c $. Obviously, although the
prediction is material independent, its formulation in terms of a
reduced temperature is not.

Note that the dimensionality of $\ga$ here is changed with respect
to the free energy potential of Eq. (\ref{3pot}), through a
convenient $m_e$ factor -- this is because the dimension of the
scalar field in G-L theory is $[\Ph^2]=L^{-3}$, its square
representing a density, while in field theory $[\Ph]=L^{-1}$. The
electron mass determines the conversion as it is absent from the
kinetic term of the Lagrangean density of field theory, $
\partial_\mu\Ph \partial^\mu\Ph $, while present in the
corresponding condensed matter free energy term, $ (\hbar^2 / 2
m_e) \nabla^2\Ph$ (or, equivalently, in the coherence length:
$\xi^2_{FT} = 1/ \al'$ \textit{vs.} $\xi^2_{cm} = \hbar^2 / 2 m_e
\al'$).

Table \ref{table2} also includes the quantities $\al'$ and $\be$,
both in SI and natural units. As explained above, conversion is
not direct, but achieved through the multiplicative factor $m_e$.

The results in Table \ref{table2} include the cubic term predicted
by Ref. \cite{Halperin}. In the absence of an applied magnetic
field, each momentum-fluctuation of the gauge field $\vec{A}$ has
an expectation value given by the equipartition theorem. When
suitably integrated over the momentum space (with a cutoff $\La$
of the order of $\xi_0^{-1}$),

\beq 
\langle A^2\rangle _\Ph = 4 {\mu_0 \over \pi} \La T_c - 8
\mu_0 {e \over \hbar c} \sqrt{\pi \mu_0} T_c |\Ph|~~.
\label{meangauge} 
\eeq

Since $A^2$ couples to $\Ph^2$ in Eq. (\ref{free}), this
translates into an unimportant correction to the scalar field
mass, plus a negative cubic term, given by $- 8 \mu_0 (e / \hbar
c)\sqrt{\pi \mu_0} T_c |\Ph|^3~$. This term implies a shift in
the superheating temperature (at zero field), of

\beq 
\De_T = 7.25 \times 10^{-12} T_c^3 H_c(0)^2 \xi_0^6
~~,
\label{DeltaT} 
\eeq

\noindent with $H_c(0)$ in \textit{Gauss} and $\xi_0$ in $\mu m$.

This shift lies beyond experimental accessibility, since it
requires a temperature sensitivity of $10^{-6} K$ (for Al;
$10^{-9} K$ for Sn). However, such an experiment \textit{might} be
performed with Al, using state of the art relative temperature measurement
techniques.

Surprisingly, for both materials the slopes of $\ga$ predicted by
TFT and Ref. \cite{Halperin} have similar magnitudes, $\sim
10^{-7} eV$. This is an indication of the analogous underlying
mechanisms behind them: the thermal averaging of the gauge field
in condensed matter can be thought of as equivalent to finite
temperature vacuum polarization in high energy physics (expressed
by the renormalization of 1-loop Feynmann diagrams).

\section{Topological defect formation}

Let us now discuss some possible implications of the inclusion of
the cubic term in the mean-field potential. Since temperature
sensitivity measurements constrain $\ga_G < 10^{-2}$, we always
assume $\ga^2 \ll \al \be $.

The K-Z mechanism predicts a density of topological defects
(vortices), $n \simeq \xi^{-2}_0({\ta_0/\ta_q})^\nu$, where $\ta_0
= \pi \hbar / 16 k_B T_c $ is the characteristic time scale, given
by the Gorkov equation, $\ta_q$ is the quench time, and $\nu$ is a
critical exponent. Moreover, it rests upon the assumption that
there is a single topological defect per $\xi^2_0$ area and,
therefore, one must look for changes induced in this quantity.
However, since the characteristic scales of the problem are
obtained via linearization of the G-L equations, close to $T_c$ and
when the order parameter is small, we see {\it no} changes in this
prediction.

On the other hand, for a thin slab of of width $L_z$, the H-R
mechanism predicts a defect density of the order $n \simeq (e/2
\pi) T^{1/2} L_z^{-1/2} \hat{\xi}^{-1}$, where $\hat{\xi} \sim 2
\pi/\hat{k}$ is the domain size immediately after the transition.
This quantity is related to the highest wavenumber $\hat{k}$ to
fall out of equilibrium and is obtained from the adiabaticity
relation

\beq 
\left\vert{d\om(k) \over dt}\right\vert = \om^{2}(k)~~,
\label{dispersion} 
\eeq 
for a given dispersion relation $\om(k)$.
In the underdamped case, $\om(k)=\sqrt{k^2+m_\ga^2}~$, with a
photon mass given by $m_\ga^2 = 2e^2 |\Ph|^2 = -2e^2 \al / \be~$.
Thus we obtain $\hat{k} \sim \sqrt[3]{\al'e^{2}/\be\ta_q}$, and
hence $n\propto\ta_q^{- 1/3}~$. Here, the introduction of a cubic
term in the potential will change the photon mass, as the true
vacuum shifts to

\beq 
\Ph = {-3\ga+\sqrt{-16\al\be+9\ga^2}\over4\be}~~.
\label{field} 
\eeq

\noindent 
However, since $\ga_G \equiv 3 \ga / 2 \sqrt{|\al|\be}
\ll 1 $, the effect of the cubic term is too small to
significantly change the H-R result.

Another effect related to metastability concerns the non-vanishing
probability of the order parameter to quantum tunnel from the the
symmetric (false) vacuum towards the non-symmetric vacuum.
Following Refs. \cite{Coleman, Linde2}, the rate of transition per
unit volume and time to the true vacua is given, in the thin wall
approximation, by

\beq 
{\Gamma \over V \Delta t} = T^4\left({S_3 \over 2
\pi~T}\right)^{3/2}e^{-S_3/T}~~, 
\label{decayingrate} 
\eeq

\noindent where

\beq 
S_3(T) = {2\pi \over 81} {1 \over \be^7 \sqrt{\be}} {\ga^9(T)
\over \ep^2(T) } 
\label{thinwall} 
\eeq 
is the Euclidean action,
and $\ep(T)$ is the ``depth" of the true vacuum.

The thin wall approximation is valid whenever the barrier's height
is much greater than $\ep$. This is true when $\ga$ is comparable
to the other parameters, namely when $\ga_G \sim 1$. For
eventually smaller values of $\ga$, like those predicted by TFT
and Ref. \cite{Halperin}, the approximation fails. In fact, we
have shown that the current temperature sensitivity of $ 10^{-3} K
$ only allows values of $ \ga_G $ smaller than $ 10^{-2}$.
Therefore, we must conclude that the barrier's height is not
comparable to the true vacuum's depth, and the field should always
tunnel through it (i.e. with a probability close to unity).
Because of this, there is no concern that defects may not have
time to nucleate within the resolution time of the measuring
device, as would happen if the potential barrier were high and
diminished too slowly.

\section{Conclusions}

In this work we have examined a possible description of a type-I
superconductive phase transition by introducing a cubic term in
the G-L mean-field potential, inspired by a gauge field thermal
averaging \cite{Halperin}, and also by analogy with TFT.

Our analysis of the bounds derived from the superheating field and
temperature constraints clearly show that the contribution of any
cubic term is small compared to other parameters in the G-L
potential. Thus the following conclusions can be drawn: 
First, the superheating temperature
shift induced by a cubic term, derived either from Ref.
\cite{Halperin} or from TFT, increases with
decreasing G-L parameter $\ka$ ($\De t_{sh}(Sn) \sim 10^{-9}$; $\De
t_{sh}(Al) \sim 10^{-6}$). This suggests that future experiments
to search for a TFT cubic term should be conducted with extreme
($\ka \ll 1$) type-I materials, for example $\al$-tungsten, with $
T_c = 15.4 \pm 0.5 ~mK $, $ H_c = 1.15 \pm 0.03 ~G $. Similarly,
the shift in the supercritical field might be reinvestigated using
a DC SQUID, which currently possess a sensitivity of $10^{-5}
\ph_0 / \sqrt{Hz} $, or $10^{-6}G $ over a 10 $\mu$m grain
diameter.

Furthermore, the impact of any cubic term on the defect density
predictions of K-Z \cite{Kibble,Zurek} or H-R
\cite{Hindmarsh-Rajantie} is negligible, with no suppression or
slowing down of defect production because the potential barrier
due to $\ga$ is not sufficient to prevent nucleation.

These considerations suggest that, all else being equal,
experiments to detect topological defect formation in type-I
superconductors would observe a  reduction of the predicted H-R
defect densities by $10-100$ depending on choice of material.
Recent calculations \cite{Shapiro} however suggest that the defect
structure formed in type-I materials survives significantly longer
than in type-II. Given that the type-I estimate is of order
$10^{-4}$ seconds, it seems possible that the disadvantage in
$\xi$ might be compensated by simple measurability.

\section{Appendix}

Including the presence of a magnetic field with a $\Ph$ dependent
magnetic moment, the conditions for the reduced superheating field
$ h_{sh} \equiv H_{sh} / H_c $ are

\vskip 0.5cm

\par (i) minimum

$$ h_{sh}^2 = {8 \over 9}\left[
{ \Ph_0^2 ~(1 + \ga_G \Ph_0 - \Ph_0^2) ~sinh^2 (x_0) \over 1 + {sinh
(2 x_0) \over 2 x_0} - {2 sinh^2 (x_0) \over x_0^2}} \right] $$

\vskip 0.5cm

\par (ii) inflexion point

$$ h_{sh}^2 = {4 \over 9} \left[{ \Ph_0^2 ~(1 + 2 \ga_G \Ph_0 - 3 \Ph_0^2)~ x_0^2
sinh^2 (x_0) \over 3 ~sinh^2 (x_0) - x_0^3 ~coth (x_0) - x_0^2 -
{1 \over 2} x_0 ~sinh(2 x_0)}\right] $$

\par

\noindent 
where $ x_0 \equiv \Ph_0 a / \la $.

Since the diameter $a$ is much larger than the penetration depth
$\la$, we can take the limit $ x_0 \rightarrow \infty $. The above
conditions become

$$ h_{sh}^2 = {8 \over 9} \Ph_0^3 ~(1 + \ga_G \Ph_0 - \Ph_0^2)~ {a \over \la} $$

\noindent and

$$ h_{sh}^2 = {4 \over 9} \Ph_0^3 ~(1 + 2 \ga_G \Ph_0 - 3 \Ph_0^2)~ {a
\over \la}~~.$$

Solving for $\Ph_0$ we get

$$ 2 (1 + \ga_G \Ph_0 - \Ph_0^2) = -(1 + 2 \ga_G \Ph_0 - 3 \Ph_0^2)
$$

\noindent 
which implies that

$$ \Ph_0 = \sqrt{{3 \over 5} + {4 \over 25} \ga_G^2 } + {2
\over 5} \ga_G \simeq \sqrt{3 \over 5} + {2 \over 5} \ga_G~~.
$$

\vskip 0.2cm

Substituting in the first expression, we obtain

$$ h_{sh}^2 = {8 \over 9}(\sqrt{3 \over 5} + {2 \over 5} \ga_G)^3
(1 + \ga_G (\sqrt{3 \over 5} + {2 \over 5} \ga_G) - (\sqrt{3 \over
5} + {2 \over 5} \ga_G)^2) {a \over \la} $$

\noindent which, to first order in $\ga_G$, becomes

$$ h_{sh} = {4 \over 5 \sqrt[4]{15}} \left(1 + {4 \over \sqrt[4]{15}} \ga_G \right)
\sqrt[4]{a \over \la}~~,$$

which is the result Eq. (\ref{h_sh_gamma}) as

$$ h_{sh}^0 = {4 \over 5 \sqrt[4]{15}}\sqrt[4]{a \over \la}~~.$$

\begin{acknowledgments}

\noindent The authors wish to thank Lu\'\i s Bettencourt, Arttu
Rajantie, Mark Hindmarsh and Boris Shapiro for helpful suggestions
and valuable insights on various aspects of this paper. The
research was partially developed while JP, TAG and PV were
participants in the COSLAB 2002 School, in Krakow; we are grateful
to the staff of the Jagiellonian University for their hospitality.
The work was supported by grants PRAXIS/10033/98 and CERN/40128/00
of the Portuguese Funda\c{c}\~{a}o para a Ci\^{e}ncia e
Tecnologia.

\end{acknowledgments}

\end{document}